\documentclass[]{aa}

\usepackage{txfonts}
\usepackage{float}
\usepackage{graphicx}
\usepackage{subfigure}
\usepackage{tabularx}
\usepackage{epsfig}

\begin{document}

\title{The merging cluster of galaxies Abell~3376: an optical
  view.  \thanks{Based on observations taken with the CTIO Blanco and
    SOAR telescopes. This research has made use of the VizieR and NED data
    bases.}}

\author{
F.~Durret\inst{1} \and
C.~Perrot\inst{1,2} \and
G.B.~Lima Neto\inst{3} \and
C.~Adami\inst{4} \and
E.~Bertin\inst{1} \and 
J.~Bagchi\inst{5}
}

\offprints{F.~Durret \email{durret@iap.fr}}

\institute{
UPMC-CNRS, UMR7095, Institut d'Astrophysique de Paris, F-75014, Paris, France 
\and
Observatoire de Paris, 75014 Paris, France 
\and
IAG, USP, R. do Mat\~ao 1226, 05508-090, S\~ao Paulo/SP, Brazil 
\and
Aix Marseille Universit\'e, CNRS, LAM (Laboratoire d'Astrophysique de
 Marseille) UMR 7326, 13388, Marseille, France 
\and
IUCAA, Pune University Campus, Post Bag 4, Pune 411007, India 
}

\date{Accepted . Received ; Draft printed: \today}

\authorrunning{Durret et al.}

\titlerunning{The cluster Abell~3376}

\abstract 
{Abell~3376 is a merging cluster of galaxies at redshift $z=0.046$.
  It is famous mostly for its giant radio arcs, and shows an elongated
  and highly substructured X-ray emission, but has not been analysed
  in detail at optical wavelengths.}
{In order to understand better the effects of the major cluster merger
  on the galaxy properties, we analyse the galaxy luminosity function
  (GLF) in the B band in several regions, as well as the dynamical
  properties of the substructures.}
{We have obtained wide field images of Abell~3376 in the B band and
  derive the GLF applying a statistical subtraction of the background
  in three regions: a circle of 0.29~deg radius (1.5~Mpc) encompassing
  the whole cluster, and two circles centered on each of the two
  brightest galaxies (BCG2, northeast, coinciding with the peak of
  X-ray emission, and BCG1, southwest) of radii 0.15~deg
  (0.775~Mpc). We also compute the GLF in the zone around BCG1, which
  is covered by the WINGS survey in the B and V bands, by selecting
  cluster members in the red sequence in a (B-V) versus V
  diagram. Finally, we discuss the dynamical characteristics of the
  cluster implied by a Serna \& Gerbal analysis.  }
{The galaxy luminosity functions (GLFs) are not well fit by a single
  Schechter function, but satisfactory fits are obtained by summing a
  Gaussian and a Schechter function.  The GLF computed by selecting
  galaxies in the red sequence in the region surrounding BCG1 can also
  be fit by a Gaussian plus a Schechter function.  An excess of
  galaxies in the brightest bins is detected in the BCG1 and BCG2
  regions.  The dynamical analysis based on the Serna \& Gerbal method
  shows the existence of a main structure of 82 galaxies which can be
  subdivided into two main substructures of 25 and 6 galaxies.  A
  smaller structure of 6 galaxies is also detected.  }
{The B band GLFs of Abell~3376 are clearly perturbed, as already found
  in other merging clusters. The dynamical properties are consistent
  with the existence of several substructures, in agreement with a
  previously published X-ray analysis.}

\keywords{galaxies: clusters: individual: Abell~3376, Galaxies:
  luminosity function}

\maketitle

\section{Introduction}
\label{sec:intro}

\begin{figure*}[!htb]
\centering
\includegraphics[width=16cm]{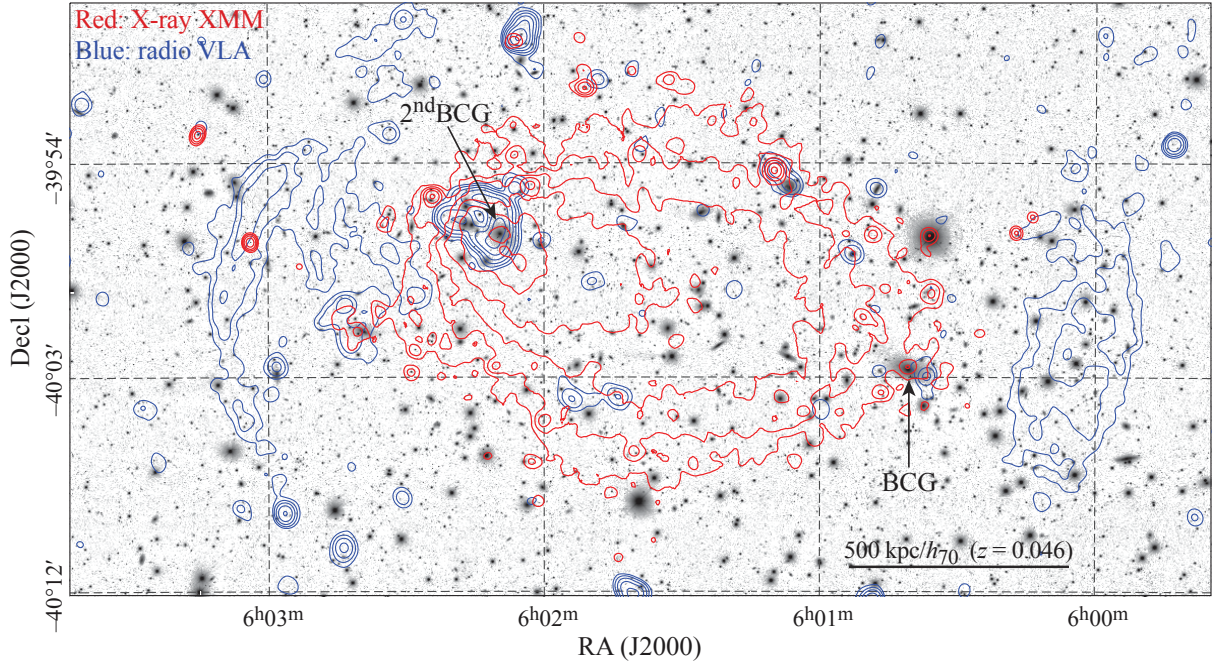}
\caption{B-band image of Abell~3376 with the X-ray (XMM-Newton) and
  radio (VLA, 1.4~GHz) emissions superimposed in red and blue
  respectively. The positions of the two BCGs are indicated, as well
  as the scale. North is up and east to the left. }
\label{fig:image}
\end{figure*}

The detailed analysis of clusters of galaxies at optical wavelengths
allows to investigate their galaxy content and distribution, through
the study of the galaxy luminosity function (GLF).  Numerous papers
have been published on galaxy luminosity functions in ``normal''
clusters of galaxies as well as in merging clusters.  The pioneering
works by Rood (1969), Peebles (1969), Rood \& Abell (1973), Schechter
(1976), or Dressler (1978), were later followed by many others at all
wavelengths. More and more distant clusters were also analysed,
reaching redshifts z=0.3 (Andreon 2001), z=0.64 (Massarotti et
al. 2003), and z$\sim 1.2$ (Drory et al. 2003).  GLFs are useful to
characterize the relative distributions of bright and faint galaxies
in various regions of clusters, and their shapes can be affected by
merging events: the GLFs of several merging clusters have been found
to show dips and wiggles, as well as an excess of bright galaxies (see
e.g. Durret et al. 2011 and references therein). Besides, the faint
end slopes of GLFs can be influenced by the infall of galaxies along
filaments feeding clusters (e.g. Adami et al. 2009).

When a sufficient number of spectroscopic galaxy redshifts is
available, it becomes possible to derive the dynamical properties and
estimate the mass of the cluster, as well as to characterize
substructures, if any.  X-ray data, in particular temperature and
metallicity maps of the X-ray gas are a natural complement to search
for substructures. Coupled with numerical simulations, they can allow
for a given cluster to draw a picture of the history of its formation (see
e.g. Durret et al. 2011 and references therein).

Abell 3376 is a merging cluster of galaxies at redshift z = 0.046
(distance modulus 36.40).  Its most remarkable feature is the
existence of giant ($\sim 2 \times 1.6$~Mpc) ring-shaped non-thermal
radio emitting structures (``radio relics'') discovered by Bagchi et
al. (2006, see their Fig. 1). A recent polarization and radio spectral
study of these giant radio relics in Abell~3376 using the Giant
Metrewave Radio Telescope (GMRT) and the Very Large Array (VLA)
suggested that diffusive shock acceleration of particles in low Mach
number (M$\sim 2-4$) shocks produced in energetic cluster merger
events are responsible for the synchrotron radio emission (Kale et
al. 2012). The large distance ($>900$~kpc) between the position of the
BCG (BCG1) and that of the X-ray peak is also remarkable and indicative
of a major merger.

The fact that Abell~3376 is a merging structure is  confirmed in
X-rays. Its XMM-Newton image is strongly elongated along
the northeast-southwest axis joining the two giant radio arcs, and the
temperature and metallicity maps of the X-ray gas show strong
inhomogeneities (see Bagchi et al. 2006, Fig.~2). Recent numerical
simulations by Machado \& Lima Neto (2013) based on the parallel SPH
code Gadget-2 have been able to reproduce the X-ray emissivity map,
and suggest an approximately head-on collision with a mass ratio of
about 3:1, observed about 0.2~Gyr after the instant of central
passage, and taking place very close to the plane of the sky.  Still
another proof for merging resides in the fact that the brightest
cluster galaxy (hereafter BCG1) is far from the region with strong
X-ray emission, as seen in Fig.~\ref{fig:image}: its coordinates are
$90.17125^o, -40.04444^o$, while the brightest galaxy close to the
X-ray peak (BCG2) is located at $90.54041^o, -39.95000^o$ (J2000.0).
Note that the direction joining the two BCGs roughly coincides with
the direction joining the two giant radio arcs.

Among the few optical studies performed on Abell~3376, we can note
that of Escalera et al. (1994) who detected the presence of several
substructures in the galaxy distribution.  Abell~3376 is part of the
WINGS survey (Fasano et al. 2006, Varela et al. 2009), but the
WINGS catalogue in the B and V bands is centered on BCG1 and
does not cover the entire cluster (see Fig.~\ref{fig:cercles}). The
existence of substructures was confirmed by Ramella et
al. (2007). However, no detailed optical analysis of the entire
cluster is available.  We will derive here the B band GLF for the
overall cluster, as well as for two circular regions centered on each
of the BCGs (see Fig.~\ref{fig:cercles}).

We will also discuss the dynamical properties of Abell~3376 based on
the redshifts and magnitudes available for cluster galaxies, to which
we apply the Serna \& Gerbal (1996) method. In this way we will see
that it is possible to separate and characterize, both spatially and
in redshift, several dynamically distinct substructures.

The paper is organized as follows: the observations and data reduction
are described in Section~\ref{sec:data}, the GLFs computed in three
regions are presented in Section~\ref{sec:GLF}, the dynamical analysis
is discussed in Section~\ref{sec:dyn} and a brief discussion of our
results together with the main conclusions are given in
Section~\ref{sec:concl}.

\section{Observations and data reduction}
\label{sec:data}

\subsection{Optical data}

We have obtained optical images in the B band with the CTIO 4m
telescope and the MOSAIC2 camera (scale 0.266~arcsec/pixel) on the
night of October 12--13, 2010.  Four images (to allow dithering) were
taken for each of two adjacent regions covering the east and west
parts of the cluster (more or less coinciding with the two BCGs), with
individual exposure times of 300~s.

Since the cluster was observed at the beginning and at the end of the
night, the airmass variations were important from one frame to
another, so the images were corrected individually for airmass after
the usual bias and flat field corrections. They were then assembled
into two images (the east and west images), using the SCAMP (Bertin
2006) and SWarp (Bertin et al. 2002) software packages. Since the
seeing varied between the east (1.26 to 1.32~arcsec) and west (1.46 to
1.59~arcsec) images, the two images were analysed separately, and the
corresponding catalogues were combined later. For the objects located
in the small zone common to both images, we chose to use the
measurements coming from the east image for which the seeing was
better.

A large image was  obtained to analyse the overall structure of the
cluster at various wavelengths, as displayed in Fig.~\ref{fig:image}.
This large image covers 1.16065~deg~$\times$~0.5947~deg = 0.690238~deg$^2$.

In order to obtain a photometric calibration of these images, since no
standard stars had been taken during the night of the observation, we
reobserved the central region of Abell~3376 the night of September
27--28, 2011 with the SOAR telescope and the SOI camera in the same
band. This image was taken in photometric conditions with an exposure
time of 600~s, and covered a region of 0.084125~deg $\times$
0.08075~deg = 0.0067931~deg$^2$ (scale 0.15~arcsec/pixel) centered on
BCG1. It was calibrated photometrically with Landolt standard
stars. We then cross-identified objects present in the Blanco and SOAR
images, and calibrated the Blanco image.  

\begin{figure}[!htb]
\centering
\includegraphics[width=7.0cm]{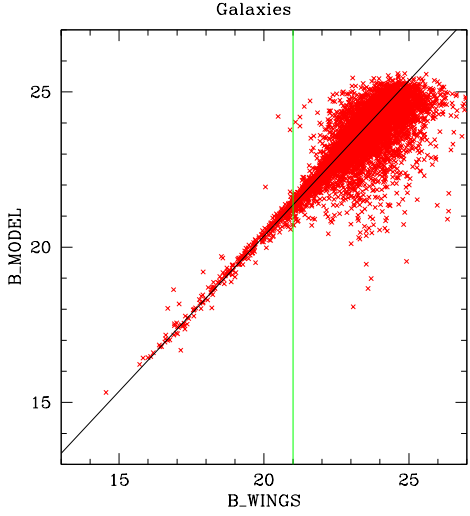}
\caption{B magnitudes of our initial catalogue as a function of B
  magnitudes from the WINGS catalogue, for objects classified as
  galaxies by WINGS. The black line shows the shift of 0.36 between
  our magnitudes ({\tt MAG\_MODEL} from SExtractor, see Section 2.2) and
  the WINGS magnitudes, estimated for objects brighter than B=21
  (green line). }
\label{fig:Bus_BWINGS}
\end{figure}

In order to test this calibration, we retrieved from
VizieR\footnote{http://vizier.u-strasbg.fr/viz-bin/VizieR} the
catalogue from the WINGS survey in the B and V bands (Varela et
al. 2009). Though the WINGS image does not encompass the full cluster,
it covers a much larger area than our SOAR image, and should therefore
provide a more reliable photometric calibration. We cross--identified
our B band catalogue with the WINGS catalogue separately for objects
classified as galaxies and as stars by WINGS. The shifts found between
our magnitudes and the B magnitudes of WINGS, for B$<21$ (to avoid
faint objects that have larger magnitude errors) were: B-BWINGS = 0.50
(dispersion 0.07) for stars and B-BWINGS = 0.36 (dispersion 0.20) for
galaxies. Though the quality of the calibration based on stars is
expected to be better, if we apply this shift to our B magnitudes the
histograms of the (B-V) for galaxies values appear quite different for
our B magnitudes and for the WINGS B magnitudes. On the other hand,
these histograms are fully consistent if we apply B-BWINGS = 0.36 (see
Fig.~\ref{fig:bmoinsv}). So in the following we will apply this
correction (illustrated in Fig.~\ref{fig:Bus_BWINGS}) to our B band
magnitudes.

We now concentrate only on the galaxies of the deep large B band Blanco image.

\subsection{SExtractor analysis and star--galaxy separation}

\begin{figure}[!htb]
\centering
\includegraphics[width=7.0cm]{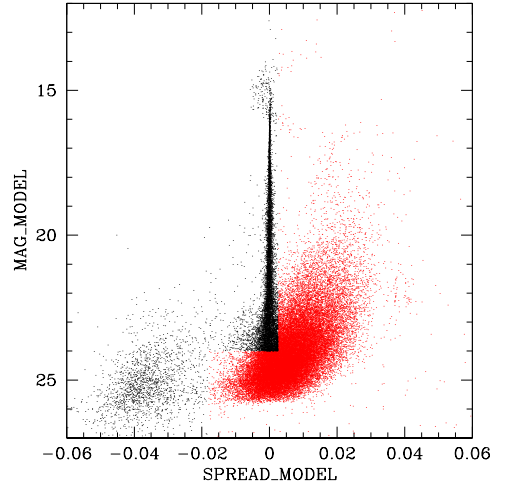}
\caption{Model magnitudes as a function of the {\tt SPREAD\_MODEL} parameter.
Sources classified as galaxies are highlighted in red.}
\label{fig:galstarsep}
\end{figure}

All sources in the eastern and western images were extracted independently using 
the SExtractor software (Bertin \& Arnouts 1996), and the resulting catalogues 
merged to remove duplicates. Before extraction, models of the Point Spread 
Function (PSF) for both images were derived with the PSFEx tool (Bertin 2011) 
from non-saturated point sources with a signal-to-noise ratio (as measured by 
SExtractor's {\tt SNR\_WIN} parameter) higher than 20. Spatial variations of the 
PSF were modelled as a third degree polynomial of the pixel coordinates. We set
the reference aperture for normalising the PSF at 20 pixels (5.3~arcsec).

SExtractor was run in single-image mode, using the experimental two-dimensional
galaxy model-fitting feature (Bertin 2011). The fitted model is the sum of a
de~Vaucouleurs ``bulge'' and an exponential ``disk'' components, centred on the
same position, and convolved with the local PSF model. Our photometric
measurements are based on SExtractor's {\tt MAG\_MODEL} ``asymptotic'' model
magnitudes, which correspond to the  integral of the best-fitting bulge+disk
model extrapolated to infinite radius.

The catalogues obtained for the east and west images were then merged,
giving a catalogue of 45538 objects, after objects in common (1725)
were eliminated.

Star-galaxy separation was performed based on SExtractor's {\tt
  SPREAD\_MODEL} estimator (e.g., Desai et al. 2012, Bouy et
al. 2013).  Fig.~\ref{fig:galstarsep} shows the distribution of
detections from both eastern and western fields in a {\tt MAG\_MODEL}
{\it vs} {\tt SPREAD\_MODEL} diagram; the point-source locus centred
around {\tt SPREAD\_MODEL}$\approx 0$ and the clump from faint
residual cosmic rays at {\tt SPREAD\_MODEL}$\approx -0.04$ are well
visible. Based on a visual inspection of the diagram, we classified as
galaxies all detections with {\tt SPREAD\_MODEL}$>-0.018$ and fainter
than B=24, or with {\tt SPREAD\_MODEL}$>0.0025$.

In all fields, we have checked by eye all the objects brighter than
B$\sim 17$, and eliminated a few bright stars (${\rm 15<B<17}$)
wrongly classified as galaxies.

We checked that the star counts as a function of magnitude agreed with
those of the Besan\c con model for the Galaxy (Robin et al. 2003),
thus confirming the validity of our star-galaxy classification.

Since BCG1 was saturated in our image, we took for this galaxy the B
magnitude given by the WINGS catalogue: B=14.55. For BCG2, which was
also saturated on our image, we took a somewhat arbitrary magnitude
B=15.00. The choice of this value was motivated by the fact that
  on the image BCG2 is smaller than BCG1, suggesting it is fainter
  than BCG1, and BCG2 cannot be much fainter than 15 because otherwise
  it wouldn't saturate.  

We ended up with a catalogue of 34660 galaxies, on which our analysis
will be based. This catalogue will be made available electronically in
VizieR.

\subsection{Completeness}

\begin{figure}[!htb]
\centering
\includegraphics[width=7.0cm]{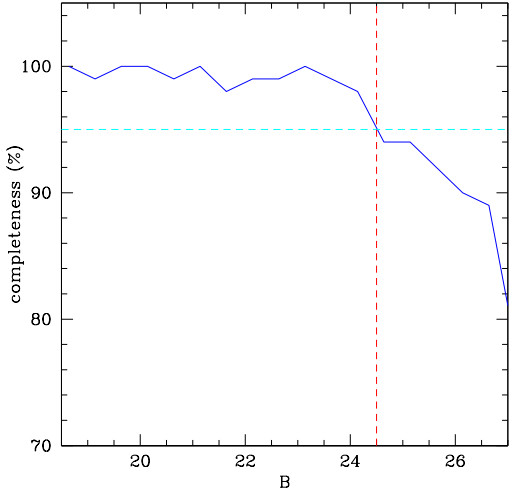}
\caption{Completeness as a function of magnitude estimated for stars
  in our image (see text): 95\% completeness (cyan dashed line) is
  obtained for B$\leq 24.5$ (red dashed line). This translates to 95\%
  completeness at B$\leq$24. for galaxies. }
\label{fig:completude}
\end{figure}

In order to check down to which magnitude we could reasonably compute
and fit the GLF, we performed simple simulations to estimate the
completeness of our galaxy catalogue as a function of magnitude. Our
method is to add ``artificial stars'' (i.e. 2D Gaussian profiles with
the same full-width-at-half-maximum as the average image point spread
function) of different magnitudes to the CCD images and attempt to
recover them by running SExtractor again with the same parameters 
for object detection and classification as on the original images. In
this way, the completeness was measured on the original images.

These simulations give a completeness percentage for stars of
  about $\sim$95\% for B$\leq 24.5$ (see Fig.~\ref{fig:completude}).
  This is obviously an upper limit for the completeness level for
  galaxies, because stars are easier to detect than galaxies. However,
  we have shown that this method yields a good estimate of the
  completeness for normal galaxies if we apply a shift of $\sim
  0.5$~mag (see e.g. Adami et al. 2006). We will hereafter consider
  that our galaxy catalogue in the B band is $\sim$95\% complete for
  B$\leq 24$. For galaxies belonging to Abell~3376 this corresponds to
  an absolute magnitude of ${\rm M_B \sim -12.5}$.

\subsection{Magnitude distribution and cross--check}

\begin{figure}[!htb]
\centering
\includegraphics[width=7.0cm]{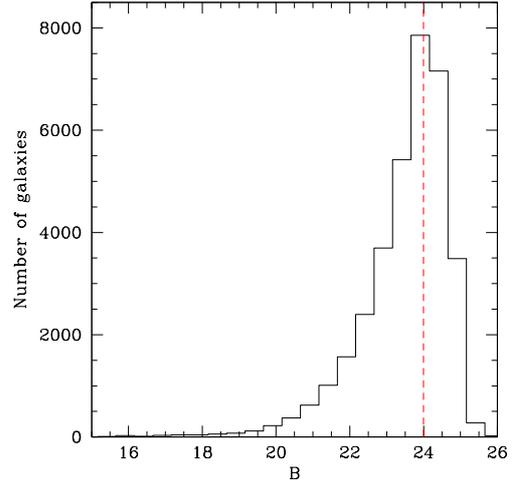}
\caption{Galaxy B band magnitude histogram of the full image. The
  vertical red dashed line at B=24 shows the 95\% completeness estimated
for galaxies.}
\label{fig:histomag}
\end{figure}

The galaxy B band magnitude histogram obtained for the full image is
shown in Fig.~\ref{fig:histomag}. The decrease in galaxy counts for
B$>24$ indicates that the completeness limit of 95\% at B=24.
estimated in the previous section could even be a little optimistic,
so we will not push our analysis at magnitudes fainter than B=23.5
(absolute magnitude $\sim -13$ at the cluster redshift).

\begin{figure}[!htb]
\centering
\includegraphics[width=7.0cm]{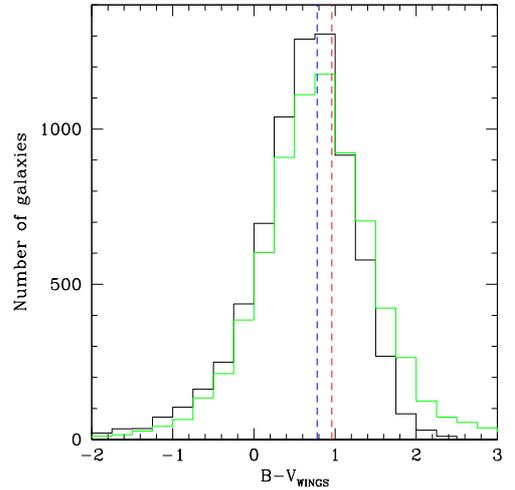}
\caption{Histogram of the (B--V) colour computed with our B band data
  (black line) and with the B band WINGS catalogue (green line), with
  V magnitudes taken from WINGS. Average colours for an Sab and an
  elliptical galaxy (taken from Fukugita et al. 1995) are indicated in
  blue and red respectively.} 
\label{fig:bmoinsv}
\end{figure}

In order to test the quality of our photometric calibration (also see
Section~2.1), we computed the histogram of the (B-V$_{\rm WINGS}$) colours,
both for our B band data and for the WINGS B band data. The result is
displayed in Fig.~\ref{fig:bmoinsv} and shows that both histograms
agree, and that the colour histograms are consistent with those of
normal Elliptical and Sab galaxies (respectively 0.96 and 0.78 at z=0,
taken from Fukugita et al. 1995).

\subsection{Galaxy redshifts}

\begin{figure}[!htb]
\centering
\includegraphics[width=7.0cm]{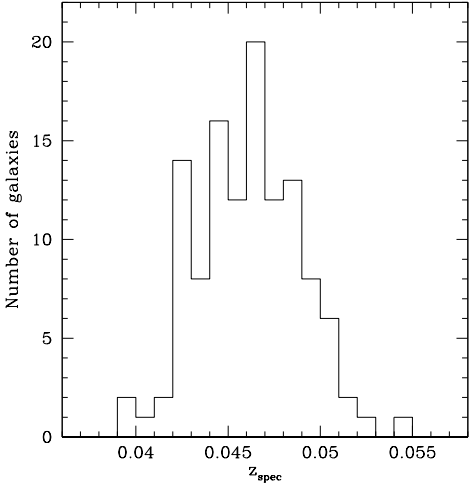}
\caption{Histogram of the 120 galaxy redshifts in the cluster range. }
\label{fig:histoz}
\end{figure}

We retrieved all the galaxy redshifts available in
NED\footnote{http://ned.ipac.caltech.edu/} corresponding to our image
(in majority coming from Cava et al. 2009). We found 213 redshifts,
out of which 120 are in the interval [0.039,0.055] that can in first
approximation be considered as corresponding to the cluster. The
redshift histogram is displayed in Fig.~\ref{fig:histoz}.  The
  mean redshift is 0.04608, corresponding to a mean velocity
  cz=13824~km/s, and to a velocity dispersion of 848~km/s.  The
  corresponding biweight quantities are 0.04604 for the mean redshift,
  cz=13813~km/s for the mean velocity, and 862~km/s for the velocity
  dispersion. The rather high value of the velocity dispersion is
expected, since the cluster is in a merging stage, and agrees with the
fact that several dynamical components are present (see
Section~\ref{sec:dyn}).  Our values agree with those obtained by Cava
et al. (2009): z=0.0461 and $\sigma_{cz}=841\pm 56$~km/s.

\section{ The galaxy luminosity function (GLF) in the B band}
\label{sec:GLF}

\subsection{Definition of three spatial regions}

\begin{figure*}[!htb]
\includegraphics[width=18.0cm,clip=true]{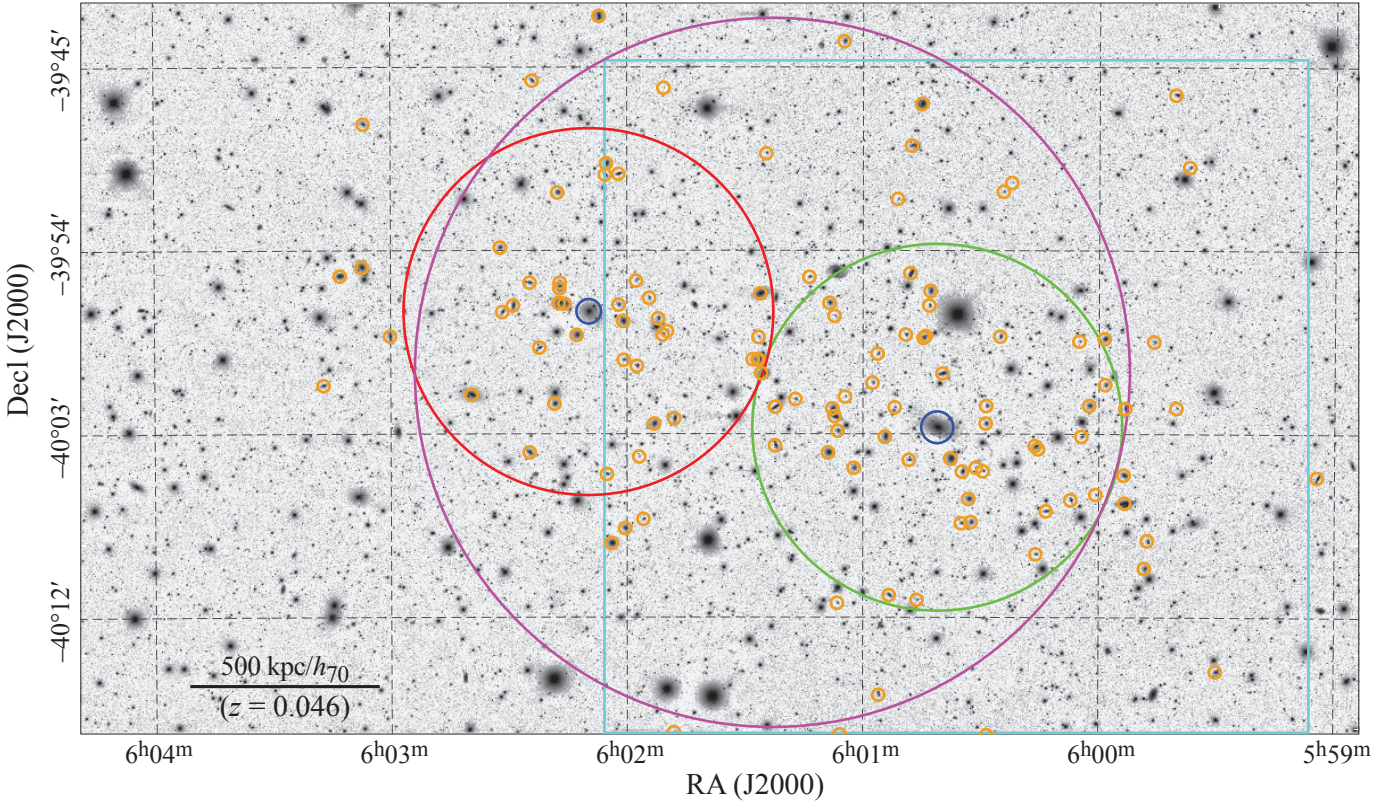}
\caption{B band image of Abell~3376. The three large circles are those
  where the GLF was extracted: the whole cluster in the magenta
    circle, the BCG1 region in the green circle and the BCG2 region in
    the red circle (see text). The two brightest galaxies are
  indicated with blue circles (BCG1, considered as the cluster center,
  is to the west, and BCG2 corresponds to the zone of maximum X-ray
  emission). The galaxies with spectroscopic redshifts in the cluster
  range are marked in orange. The cyan rectangle shows the field
  covered by the WINGS catalogue in the B and V bands. The total
  image size is $1.03 \times 0.59$~deg$^2$. North is up and east to
  the left. }
\label{fig:cercles}
\end{figure*}

Abell~3376 is characterised by a very large distance between the
position of the brightest cluster galaxy (hereafter BCG1) and that of
the second brightest galaxy (BCG2) located in the area of the X-ray
peak. This is illustrated in Fig.~\ref{fig:image}. We therefore
divided the overall image in three zones to analyse the GLF: a circle
of 0.29~deg radius (1.5~Mpc) encompassing the whole cluster, centered
on the position: 90.346~deg, $-40.0$~deg (the ``whole cluster'', coded
in magenta in several figures, with 9887 galaxies), and two circles of
radii 0.15~deg (0.775~Mpc) centered on each of the two brightest
galaxies: BCG1 at 90.17125~deg, $-40.04444$~deg (coded in green, 1969
galaxies) and BCG2 at 90.54041~deg, $-39.9500$~deg (coded in red, 3567
galaxies). The positions of these circles, together with those of the
two BCGs are shown in Fig.~\ref{fig:cercles}.

We first derive galaxy luminosity functions in the three regions
described above, obtained by applying a statistical subtraction of the
background.  For the BCG1 region, which is the only one to be
fully covered by the WINGS survey, we will also derive the GLF by
selecting cluster members from the red sequence in a (B-V) versus V
diagram. 

We estimated the numbers of objects that we could be missing due to
the presence of bright stars or to the haloes of the BCGs. We find
that the surfaces covered by bright stars correspond to 3\% to 5\% of
the total area in the BCG1 region and about 2\% in the BCG2 region.

The objects in the halos of the BCGs should normally be detected and
deblended from the BCG by SExtractor.  We checked by eye the areas
around the BCGs for galaxies undeblended by SExtractor.  All the
galaxies visible on the image located more than 1 arcmin from the BCGs
are in our catalogue. Within circles of 1~arcmin radii from the BCGs,
we found that the fraction of lost sources was about 25\%-30\%. The
ratio of the surface covered by a 1~arcmin radius circle to the total
surface covered by the BCG1 and BCG2 regions (radius 0.15 deg) is
1.2\%, so a loss of 30\% of the galaxies in a zone covering only 1.2\%
of the total area of these zones can be considered as negligible.

Since the corrections for incompleteness due to non-detections in the
haloes of bright stars or of the BCGs are small and probably not very
accurate, we decided not to apply them to our galaxy counts and GLF
computations.

\subsection{The B band galaxy luminosity functions in three regions
(statistical background subtraction)}

In order to compute the B band galaxy luminosity functions, we counted
all the galaxies in magnitude bins of 0.50~mag and subtracted
statistically the contribution of background galaxies, using the field
galaxy counts per square degree estimated by McCracken et al. (2003)
in the B filter. Since these authors give field galaxy counts for
${\rm B\geq 18}$, for galaxies brighter than B=18 we only took into
account galaxies with a spectroscopic redshift in the cluster. This
approach is justified by the fact that the completeness of the
spectroscopic data is high: the ratios of the numbers of galaxies
brighter than B=18 with measured redshifts to the total numbers of
galaxies brighter than B=18 are 78\% (42/54), 87\% (13/15) and 87\%
(20/23) in the overall cluster, BCG1, and BCG2 regions
respectively. In view of the relatively small numbers of galaxies, we
did not apply a completeness correction to the bright end of the GLF
(since these numbers would have further been distributed in various
magnitude bins) to avoid adding noise to the data. We can note that
the excesses observed in somes GLFs at very bright magnitudes are thus
lower limits.

Since the McCracken counts are given in AB magnitudes, we had to
convert our magnitudes to the AB system, applying: ${\rm B_{AB} =
  B_{Vega}-0.09}$. We also applied to our galaxy catalogue an
extinction correction of 0.186~mag, as given by NED from Schlafly \&
Finkbeiner (2011), and normalized all surfaces to 1~deg$^2$.  In view
of the small redshift of the cluster, we did not apply a K$-$correction.

\begin{figure}[!htb]
\includegraphics[width=9.0cm]{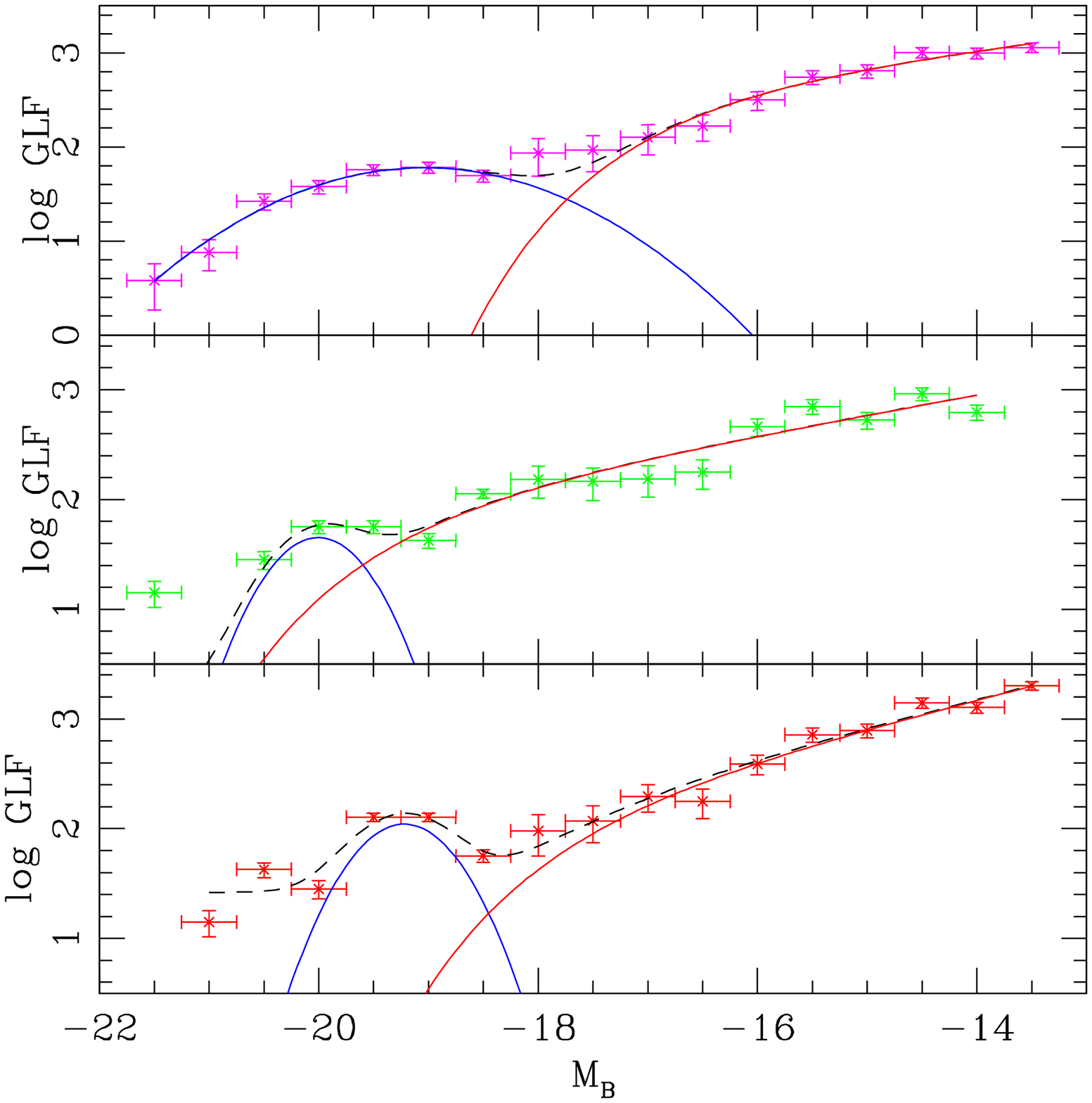}
\caption{Galaxy luminosity functions in the B band in the three zones
  of Abell~3376 shown in Fig.~\ref{fig:cercles}, with the same colour
  codes. The units are galaxy counts in bins of 0.5~magnitude and per
  square degree. The best fits obtained 
  are shown as curves: a Gaussian (in blue) at the bright end, a
  Schechter function (in red) at the faint end, and the sum of both
  shown as a black dashed line. 
}
\label{fig:GLF}
\end{figure}

\begin{table*}
  \caption{Galaxy luminosity function parameters. SS stands for statistical background 
subtraction and RS indicates red sequence selection.}
\begin{tabular}{lcrrrrrr}
\hline
\hline
Region       & Range (${\rm M_B}$) & $\Phi ^*$ & $M^*$& $\alpha$ & $A$ & $M_c$ & $fwhm$ \\
\hline
Whole cluster (SS)&[$-21.5,-13.5$]&$486\pm 110$&$-16.76\pm 0.22$&$-1.36\pm 0.06$&$60\pm 2$&$-19.04\pm 0.08$&$2.46\pm 0.11$\\
BCG1 (SS)    &[$-21.5,-14.0$]&$103\pm 14$~&$-19.38\pm 0.13$&$-1.45\pm 0.02$&$45\pm 3$&$-20.00\pm 0.03$&$0.89\pm 0.04$ \\
BCG1 (RS)    &[$-21.5,-12.5$]&$679\pm 111$&$-15.80\pm 0.18$&$-1.32\pm 0.04$&$108\pm 7$&$-18.09\pm 0.14$&$2.74\pm 0.17$ \\
BCG2 (SS)    &[$-21.0,-13.5$]&$190\pm 48$~&$-17.74\pm 0.24$&$-1.63\pm 0.03$&$110\pm 4$&$-19.2\pm0.02$&$0.94\pm 0.04$  \\
\hline
\end{tabular}
\label{tab:GLF}
\end{table*}

The resulting galaxy luminosity functions were computed as a function
of absolute magnitude, assuming a distance modulus of 36.40, in order
to make the comparison possible with other clusters. They are shown in
Fig.~\ref{fig:GLF}. The GLF of the entire cluster (in magenta) is
rather smooth, but shows an excess at ${\rm M_B \sim -18}$.  The GLF
in the region of BCG1 (in green) is quite irregular, with several dips
and wiggles and an excess in the brightest bin (${\rm
  M_B=-21.5}$). The GLF in the region of BCG2 (in red), which
coincides with the X-ray peak, is also somewhat irregular at bright
magnitudes (i.e. for ${\rm M_B < -18.5}$), with an excess at ${\rm M_B
  \leq -20.5}$ in the two brightest bins.

The fits of all three GLFs by a simple Schechter function are not good
(i.e. the fit obtained with the MIGRAD minimization of the MINUIT
software does not converge). We therefore fit them with the sum of a
Gaussian and a Schechter functions, keeping in mind the fact that a
fit including 7 free parameters (the 7th free parameter being a
constant, found to be zero in the overall and BCG1 regions, but
non-zero in the BCG2 region) and less than 20 data points could be
somewhat {\it ad hoc}.

For the Gaussian  function, we took:
$$ G(M) = A\ exp[ (-4 * ln(2) * (M-M_c)^2 ) / (fwhm^2) ] $$
where $M_c$ is the central magnitude, $fwhm$ is the full width at half
maximum and $A$ is the amplitude for $M=M_c$.

For the Schechter function, we used:
$$ S(M) = 0.4 \, \ln 10 \, \Phi^{\ast} \, y^{\alpha+1} \, e^{-y} $$
with 
$y=10^{0.4 \, (M^{\ast}-M)} $.

The fits are shown in Fig.~\ref{fig:GLF} and the parameters of the GLF
fits are given in Table~\ref{tab:GLF} (see discussion in Section~5).
We can note that the 7th parameter is not zero for the BCG2 region,
implying the existence of an excess of very bright galaxies in this
zone.

\subsection{The B band galaxy luminosity function in region BCG1
(red sequence galaxy selection)}

\begin{figure}[!htb]
\includegraphics[width=7.0cm]{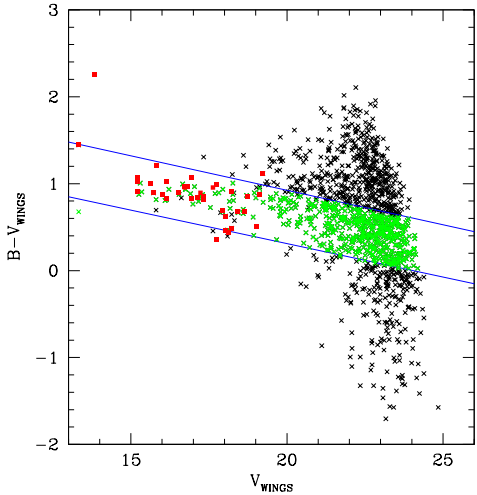}
\caption{(B-V) versus V colour-magnitude diagram for the BCG1
  region of Abell~3376. Black crosses show all the galaxies, red
  squares show the galaxies with spectroscopic redshifts in the
  cluster and green squares show all the galaxies selected as cluster
  members within the red sequence delimited by the two blue lines (see
  text). The (few) black crosses within the red sequence have spectroscopic
  redshifts outside the cluster and thus were not taken into account to
  compute the GLF.}
\label{fig:RS_green}
\end{figure}

\begin{figure}[!htb]
\includegraphics[width=7.0cm]{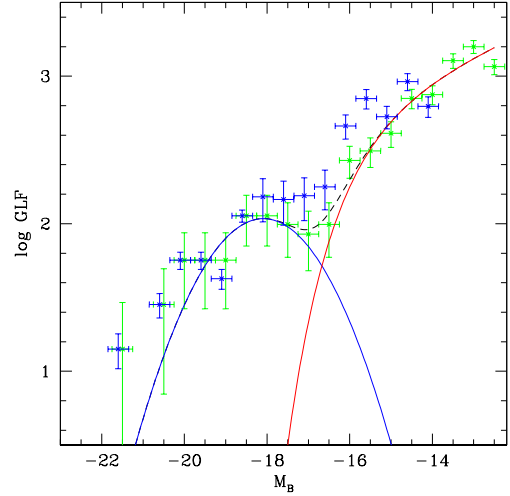}
\caption{Galaxy luminosity function in the B band for the BCG1
  region of Abell~3376, obtained by selecting galaxies along the red
  sequence in a (B-V) versus V colour-magnitude diagram (green
  points). The units are galaxy counts in bins of 0.5~magnitude and
  per square degree.  The best fit is obtained by summing a Gaussian
  (in blue) at the bright end, and a Schechter function (in red) at
  the faint end; their sum is shown as a black dashed line. The
    GLF obtained by statistical subtraction in the same region is
    shown as blue points, displaced by $-0.1$~mag to avoid
    overlapping.} 
\label{fig:GLF_green}
\end{figure}

Since the WINGS catalogue in the V band covers completely the BCG1
region, we also derived the GLF in this zone by selecting cluster
galaxies from the red sequence in the (B-V) versus V colour-magnitude
diagram, as shown in Fig.~\ref{fig:RS_green}. One can see that the red
sequence is well defined (in particular by spectroscopically confirmed
cluster members).  For V$\leq 19$ the best fit to the red sequence is:
$${\rm (B-V) = -0.077 \times V + 2.15.}$$ We select as cluster
members all the galaxies within $\pm 0.3$ on either side of this
relation, to which we add the five spectroscopically confirmed cluster
members that lie outside this zone, and BCG1. We exclude the
eight galaxies falling in the red sequence interval but with
spectroscopic redshifts outside the cluster.

The GLF obtained by this selection is displayed in
Fig.~\ref{fig:GLF_green}. Here also, a single Schechter function
cannot fit the data, but a good fit is obtained by summing a Gaussian
and a Schechter function (see parameters in Table~\ref{tab:GLF}). We
can note again an excess of galaxies in the brightest bin in this region.

We can see in Fig.~\ref{fig:GLF_green} that the GLFs obtained with
both methods for the BCG1 region agree within error bars in almost all
magnitude bins. The fact that the GLF points in the individual bins
tend to be higher in the statistical subtraction case than in the red
sequence selection could be due to the fact that the gaps between the
CCDs in the WINGS data reduce by a small amount the number of galaxies
measured in the V band, and therefore the number of galaxies
cross-identified with our B band catalogue. 

Although the GLFs obtained with the two methods roughly agree, we can
see in Table~1 that the parameters of the best fits obtained with both
methods (see Table~1) are quite different for the bright
  component (the Gaussian), while the faint end slopes do not strongly
  differ. It is rather surprising that the GLF fits for two sets of
  data that seem consistent within error bars can be so different.
  To our opinion, this illustrates the difficulty of obtaining robust
  luminosity functions and to fit them in a reliable way, and this
  should be a warning not to overinterpret data.

Assuming that all the galaxies selected in the red sequence belong to
the cluster is certainly an overestimate, although the fact that the
GLF derived from the red sequence agrees with the one computed by
statistical background subtraction tends to indicate that the
contamination is probably not too large.

We can estimate this contamination by considering the Coma cluster,
which has extensively been covered spectroscopically by the Sloan
Digital Sky Survey, which is complete to $r'=17.5$. By applying the
colour correction from Fukugita et al. (1995) for a typical elliptical
galaxy at the redshift of Abell~3376, this corresponds to ${\rm M_B
  \sim -16}$. If we count the numbers of galaxies outside the Coma
cluster (i.e. with redshifts $z<0.012$ or $z>0.035$) falling within an
interval of $\pm 0.3$ from the red sequence, we find a contamination
of 9\%. Therefore for Abell~3376 the contamination of the red
sequence that we considered by field galaxies is about 9\% for ${\rm
  M_B \sim -16}$. For fainter magnitudes (${\rm -15.5 \leq M_B \leq
  -14}$), even our Coma cluster spectroscopic catalogue is not
sufficiently complete to estimate the contamination by background
galaxies in the same way, but the fact that the values of the GLFs
(and of the faint end slopes) estimated for the BCG1 region with the
two methods (statistical subtraction of the background or red sequence
galaxy selection) do not strongly differ suggests that this contamination
is not very strong yet for  ${\rm   M_B \leq -14}$. 


These results will be discussed in more detail in Section~5.

\subsection{The bright to faint galaxy density ratio}

We quantified the distributions of galaxies as a function of magnitude
in a more global way, by estimating the densities of bright and faint
galaxies and the bright to faint galaxy density ratios, in the three
regions of interest.

For this, we computed the numbers of bright (absolute magnitude ${\rm
  M_B <-16}$) and faint (${\rm -16 \leq M_B \leq -14}$) galaxies
in the three regions analysed, and the ratios of these numbers, in the
three regions.  We find a bright to faint number ratio of 0.20
in the whole cluster and 0.15 in the BCG2 region. In the BCG1 region,
the bright to faint number ratio is 0.29, that is notably higher
  than in the other regions. {There are two possibilities: either the
  BCG1 region has a higher density of bright galaxies, or the BCG2
  region has a higher density of faint galaxies. This will be
  discussed in Section~5.}

\section{Dynamical analysis}
\label{sec:dyn}

\begin{figure}[!htb]
\includegraphics[width=8.0cm,height=6.0cm]{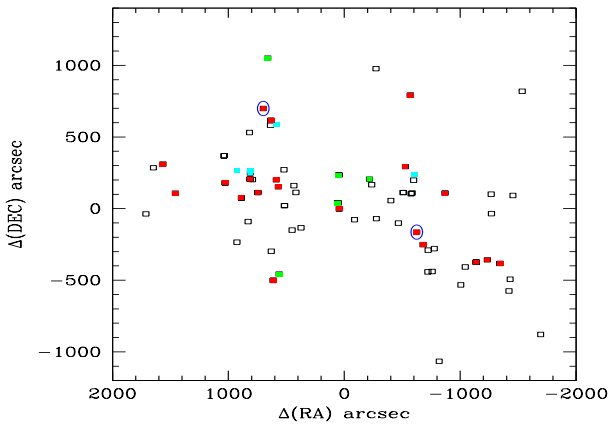}
\caption{Positions of the structures detected by the Serna \& Gerbal
  method.  The main structure (S1, 82 galaxies) is shown as black open
  squares, and includes two smaller structures, S1a and S1b, of 25 and
  6 galaxies respectively, shown in red and green. A second
  independent structure of 6 galaxies (S2) is shown in cyan. The axes
  correspond to the right ascension and declination offsets (in
  arcseconds) relative to the position of the center of the magenta
  circle (90.346~deg, $-40.0$~deg). The two blue circles show the
  positions of BCG1 (west) and BCG2 (east). }
\label{fig:SG}
\end{figure}

In order to identify the substructures present in the galaxy
distribution of Abell~3376 and to estimate their relative masses, we
applied the Serna \& Gerbal (1996) method (hereafter SG).  Very
briefly, this method calculates the potential binding energy between
pairs of galaxies and detects substructures taking into account
positions and redshifts. Here, we required a minimum number of 5
galaxies for a substructure to be present.  Assuming a value of 100
(in solar units) for the total mass to stellar luminosity ratio (but
results do not strongly depend on this quantity, see e.g. Adami et
al. 2005), galaxy magnitudes can be transformed into masses, and
approximate values can be estimated for the total (i.e. dynamical)
mass of each substructure.  Although the absolute values of these
masses are not fully reliable, mass ratios can be considered as
robust, with a typical uncertainty of the order of 25$\%$. Note
however that we neglect here the uncertainty on the M/L ratio assumed
for the galaxies. More details, as well as a discussion of the
influence of incomplete spectroscopic sampling, can be found in
Guennou et al. (2013, A\&A submitted).

We applied this method to the catalogue of galaxies with spectroscopic
redshifts and with B band magnitudes in our catalogue, including the
two BCGs (the uncertainty on their exact magnitudes should not have a
strong influence on the results).

\begin{table}
\caption{Properties of the substructures found in Abell~3376 by the SG method.
The columns are: name of the subsstructure (see text), number of galaxies,
average velocity, velocity dispersion, and mass relative to that of
the overall cluster.}
\begin{tabular}{lrrrl}
\hline
\hline
Structure & N$_{gal}$ &velocity & velocity & mass/  \\
          &          & (km/s)  & dispersion & mass (overall) \\
          &          &         & (km/s)   & \\ 
\hline
Overall   & 122      & 13595   &  1045 &  ~~1.00 \\
S1        & 82       & 13800   &   575 &  ~~0.30 \\
S1a       & 25       & 13823   &   367 &  ~~0.11 \\
S1b       & 6        & 14731   &   188 &  ~~0.004 \\
S2        & 6        & 12669   &   110 &  ~~0.002 \\
\hline
\end{tabular}
\label{tab:SG}
\end{table}

For each structure detected by the SG method, we give in
Table~\ref{tab:SG} the following parameters: number of galaxies,
average velocity, velocity dispersion, and mass relative to that of
the overall cluster.  The galaxy distributions are shown in
Fig.~\ref{fig:SG}.  The overall cluster is detected as a structure of
122 galaxies with an approximate mass of $5.3\
10^{14}$~M$_\odot$. Note however that this structure should only
  include the 120 cluster members discussed in Section~2.5, so the
  velocity dispersion of this ``Overall structure'' is obviously
  overestimated (1045~km/s instead of 862~km/s) and this is therefore
  also most probably the case for its mass. In any case, the mass of
  the ``Overall structure'' is of the same order as the virial mass
of $6.6\ 10^{14}$~M$_\odot$ estimated by Escalera et al. (1994).  The
main substructure, hereafter S1, includes 82 galaxies (34 members,
among which BCG1, and 48 probable members, among which BCG2). A much
smaller substructure (S2) of 6 galaxies is detected with an average
velocity smaller than that of the overall cluster by 920~km/s.

Structure S1 can be subdivided into two substructures, S1a and S1b,
with 25 and 6 galaxies respectively.  It is difficult to assign a
particular substructure to each of the BCGs, since all the structures
detected seem to cover a rather large area in projection on the
sky. This is in agreement with the fact that the merging takes place
in the plane of the sky, and in this case the SG method is somewhat
less efficient to discriminate subgroups.  Nevertheless, we can see
that S1a is much more massive than the other substructures, as
expected since it corresponds to the direction of the merger and to
the region where most of the brightest galaxies are concentrated.

Based on a wavelet analysis, Escalera et al. (1994) classified
Abell~3376 as ``unimodal with structures in the core''.  Their wavelet
map shows a clear elongation in the approximate direction of the
merger.  Such a unimodal classification appears rather surprising at
first sight, since there are two BCGs in the cluster. However, it
agrees with our finding that BCG1 and BCG2 are both part of S1a. It is
difficult to compare quantitatively our results to those of Escalera
et al. (1994), since they had fewer redshifts and did not give the
precise positions of their substructures, but there does not seem to
be any obvious discrepancy between their analysis and ours.

Ramella et al. (2007) searched for substructures in the area covered
by the WINGS data and detected a main structure around BCG1 (the black
dots in their figure~6) and a substructure to the northeast (the green
dots in their figure~6) which could coincide with our structure S1b.
However, since their image does not encompass the whole cluster, it is
difficult to make a more detailed comparison between our results and
theirs. 


\section{Discussion and conclusions}
\label{sec:concl}

Abell~3376 is a cluster undergoing a major merger and characterized by
a strongly elongated X-ray emission, peaked on the second brightest
galaxy BCG2, while the brightest galaxy BCG1 is located about 0.3~deg
(1.55~Mpc) south west of BCG2.

We have analysed a large image of this cluster in the B band, reaching
a completeness level of 95\% for B$\sim$24.  We computed the B band
galaxy luminosity functions (GLF) based on statistical background
subtraction in three regions: one encompassing all the cluster (the
``whole cluster''), one centered on BCG1 and the other on BCG2. The
three GLFs present excesses, dips and wiggles, and cannot be fit by a
single Schechter function. On the other hand, they can all be
satisfactorily fit with the sum of a Gaussian and a Schechter
functions. The overall cluster shows a small excess of galaxies at
${\rm M_B \sim -18}$. This may be due to the fact that we
  consider spectroscopic data for galaxies brighter than B=18 (${\rm
    M_B}\sim -18.4$) and statistical background subtraction for
fainter objects, and  the connection between the points obtained
with these two methods may not be perfect.

The GLF of region BCG1 is quite irregular and shows an excess in the
brightest magnitude bin ${\rm M_B = -21.5}$. The GLF of region BCG2
also shows some irregularities, and an excess in the two brightest
bins for ${\rm M_B \leq -20.5}$. Although we would expect the region
of BCG1 to dominate the overall cluster, we can see in Table~1 that
the best fit parameters obtained for the GLFs of the overall cluster
and of the BCG1 region are quite different, except for the faint end
slopes. The GLF best fit parameters of region BCG2 also differ from
those of the other regions, and the faint end slope is notably steeper
than in the two other zones. 


For the region around BCG1, we also computed the GLF obtained by
selecting galaxies along the red sequence in a (B-V) versus V diagram
(with V magnitudes taken from the WINGS catalogue), taking into
account the spectroscopic redshift information available.  The GLF
thus obtained is consistent within error bars with that obtained by
statistical background subtraction, except in a few magnitude bins
fainter than ${\rm M_B}\sim -16$. However, the best fit parameters of
the GLFs obtained by both methods differ quite strongly, except for
the faint end slopes (which are not fully consistent though, but we
have noticed before that the error bars computed by MINUIT were in
some cases underestimated). This illustrates the difficulty of
deriving unambiguous fits for GLFs: the error bars on each bin are
large, and changing the galaxy counts by a small amount can modify the
fit parameters obtained. Obviously, a good photometric calibration is
crucial and a check with numerous spectroscopic redshifts, going as
deep as possible in magnitude, is mandatory to ascertain the quality
of the GLFs thus obtained (the results obtained on Coma with very deep
spectroscopy by Adami et al. 2009 illustrate well this point).  This
result should be kept in mind in further studies.


As already noted e.g. in the case of Abell~1758 North, which is
clearly the result of a merger (Durret et al. 2011 and references
therein), the dips and wiggles of the GLF seem to be a characteristic
of merging clusters. The GLF of Abell~1758 North could be fit
  with a single Schechter function, so the comparison of the bright
  ends of the GLFs in Abell~3376 with this cluster are probably
  irrelevant.  However, we can compare the faint end slopes (the
  $\alpha$ parameter). The faint end slopes estimated for all the
regions of Abell~3376 are steeper than that of Abell~1758 North
($\alpha=-1.00\pm 0.02$ in the $g$ band, at a redshift of 0.279).  An
explanation could be that the mergers in these two clusters do not
have the same age and that the faint end slope varies with
time. Another possibility is that the faint end slope varies with
redshift, and is flatter at larger redshift. This is suggested by the
flat slopes determined by Rudnick et al. (2009) for clusters at
redshifts $z<0.8$.  We are in the process of analysing galaxy
luminosity functions in a sample of clusters at $0.4<z<0.9$ from the
DAFT/FADA survey in all stages of merging, from relaxed to strongly
substructured (Martinet et al. in preparation).  This should shed
light on the influence of mergers on the galaxy distributions in
clusters.

  The difference in the bright to faint number ratio in the BCG1
  and BCG2 regions can be interpreted as either due to a higher
  density of bright galaxies in the BCG1 region, or to a higher
  density of faint galaxies in the BCG2 region.

  BCG1 is usually considered as the center of the main cluster because
  it is the brightest galaxy and has a cD like morphology.  A higher
  density of bright galaxies in the BCG1 region agrees (at least
  qualitatively) with the scenario simulated by Machado \& Lima Neto
  (2013), where the subcluster around BCG2 has crossed the main
  cluster about 0.2~Gyr ago, and the cluster mass ratio is about
  3:1. In this case more bright galaxies are expected around BCG1 than
  around BCG2.

  On the other hand, the BCG2 region could have a much higher density
  of faint galaxies, and could even be more massive than the BCG1
  region. There are several arguments in favor of this hypothesis.  First,
  the number of galaxies in the BCG2 region (3567) is almost twice the
  number of galaxies in the BCG1 region (1969). Second, the number of
  bright galaxies (B$\leq 18$) is also larger in BCG2 (23) than in BCG1
  (15). Third, the faint end slope is notably steeper in region BCG2
  than in BCG1 (see Table~1). And finally, the fact that the peak of
  X-ray emission is located in BCG2 could indicate the presence of a
  deeper potential well in the BCG2 region. The only disturbing point
  with this scenario is that BCG2 appears at the edge of the
  distribution of the various structures determined by the Serna \&
  Gerbal method (see Fig.~\ref{fig:SG}). 

  We have been granted one night observing time on Blanco with DECam
  in January of 2014, to obtain very deep images of Abell~3376 to
  perform a weak lensing mass reconstruction of this cluster. This
  should allow to choose between these two scenarios.

The dynamical analysis of Abell~3376 based on the Serna \& Gerbal
(1996) method shows that the cluster (122 galaxies) contains a main
structure (82 galaxies) which can be subdivided into two substructures
S1a and S1b, of 25 and 6 galaxies respectively, the first one
containing BCG1 and BCG2. A smaller substructure of 6 galaxies is also
detected with an average velocity smaller than that of the overall
cluster by 920~km/s. S1a is much more massive than the other
substructures, as expected since it corresponds to the direction of
the merger and to the region where most of the brightest galaxies are
concentrated.  Our results agree qualitatively with those of Escalera
et al. (1994) and Ramella et al. (2007) but a quantitative comparison
is not possible. 

Abell~3376 has been studied at radio and X-ray wavelengths (Bagchi et
al. 2006), and accounted for by recent numerical simulations (Machado
\& Lima Neto 2013), but it remains poorly known in the optical, though
it was part of the WINGS survey (Fasano et al. 2006, Varela et
al. 2009). We hope to have shed some light on its optical properties
in the present paper, and plan to obtain deeper images in the future
to improve our understanding of this cluster where violent merging
events are taking place, in particular by analysing its mass
distribution through weak lensing techniques.

\begin{acknowledgements}

  We warmly thank the anonymous referee for many constructive comments that
  helped us to improve the paper.  We acknowledge long term financial
  support from CNES as well as CAPES/COFECUB program 711/11. We are
  grateful to the DAFT/FADA team for the SOAR observations of this
  cluster.

\end{acknowledgements}

\appendix

\section{Description of the electronic table}

\begin{table*}
  \caption{First ten lines of the galaxy catalogue that will be available in VizieR. 
The columns are: Galaxy number, right ascension, declination, B\_AUTO magnitude and its error, B\_MODEL magnitude and its error,  SPREAD\_MODEL and its error   }
\begin{tabular}{ccccccccc}
\hline
\hline
Number & RA (J2000.0) & DEC (J2000.0) & B\_AUTO & error & B\_MODEL & error & SPREAD\_MODEL & error \\
       & & & & (B\_AUTO) & & (B\_MODEL) & & (SPREAD\_MODEL) \\
\hline
    1 &89.6201787 &-39.7236609 &24.528  &0.179 &22.800  &0.165 & 0.026 & 0.006\\
    2 &89.6202716 &-39.7070639 &18.762  &0.004 &20.599  &1.146 & 1.000 & 1.000\\
    3 &89.6203079 &-39.7500203 &23.658  &0.135 &22.873  &1.449 & 0.011 & 0.005\\
    4 &89.6203211 &-39.7573528 &23.488  &0.120 &22.062  &0.202 & 0.018 & 0.006\\
    5 &89.6203219 &-39.7469425 &24.651  &0.209 &22.756  &1.803 & 0.019 & 0.006\\
    6 &89.6203259 &-39.7201251 &24.221  &0.143 &23.177  &4.936 & 0.019 & 0.009\\
    7 &89.6203409 &-39.6848968 &23.329  &0.128 &23.014  &0.228 & 0.020 & 0.005\\
    8 &89.6203633 &-39.7595031 &23.729  &0.122 &23.075  &0.256 & 0.008 & 0.006\\
    9 &89.6203726 &-39.7550236 &23.284  &0.109 &22.882  &2.372 & 0.019 & 0.004\\
   10 &89.6203876 &-39.7388027 &23.399  &0.111 &23.056  &0.319 & 0.007 & 0.005\\
\hline
\end{tabular}
\label{tab:galcat}
\end{table*}

The galaxy catalogue will be available in electronic form in
  VizieR\footnote{http://vizier.u-strasbg.fr/viz-bin/VizieR}. It is
  ordered in right ascension, magnitudes are in the Vega system and no
  extinction correction is applied.

  The table columns are: sequential number, right ascension in decimal
  degrees (J2000), declination in decimal degrees (J2000), B band
  MAG\_AUTO, B band error on MAG\_AUTO, B band MAG\_MODEL , B band
  error on MAG\_MODEL, SPREAD\_MODEL, error on SPREAD\_MODEL. All
  these values were computed with SExtractor (see Section~2.2).

  The galaxy catalogue was cut to B\_MODEL$\leq 25$, but obviously
  objects fainter than B\_MODEL=24 or 24.5 should be considered with
  caution.  All objects brighter than B\_MODEL$\leq 18$ were checked
  by eye in order to elliminate bright stars that had been
  misclassified as galaxies (a few tens).  The final catalogue
  includes 
  28828 objects. Note that the photometric calibration was done as
  described in Section~2.1., applying the zero point shift of 0.36
  determined from comparison with the WINGS B band catalogue for
  galaxies.  As described in Section~2.3 it is 95\% complete for
  B$\leq 24$.  The first ten lines of the catalogue are displayed in
  Table A.1.


\end{document}